\documentclass[11pt]{article}
\usepackage{graphicx}
\usepackage{amsmath}
\usepackage{hyperref}
\usepackage{geometry}
\title{Optimizing Feature Ordering in Radar Charts for Multi-Profile Comparison}
\author{Albert Dorador \\ \texttt{trustalgorithm.dev@gmail.com}}
\date{\today}

\begin{document}

\maketitle

\begin{abstract}
Radar charts  are widely used to visualize multivariate data and compare multiple profiles across features. However, the visual clarity of radar charts can be severely compromised when feature values alternate drastically in magnitude around the circle, causing areas to collapse, which misrepresents relative differences. In the present work we introduce a permutation optimization strategy that reorders features to minimize polygon ``spikiness'' across multiple profiles simultaneously. The method is combinatorial (exhaustive search) for moderate numbers of features and uses a lexicographic minimax criterion that first considers overall smoothness (mean jump) and then the largest single jump as a tie-breaker. This preserves more global information and produces visually balanced arrangements. We discuss complexity, practical bounds, and relations to existing approaches that either change the visualization (e.g., OrigamiPlot) or learn orderings (e.g., Versatile Ordering Network). An example with two profiles and $p=6$ features (before/after ordering) illustrates the qualitative improvement.
\end{abstract}

\section{Introduction}

Radar charts (also known as spiderweb or star plots) provide a compact visual summary of multivariate observations by placing feature axes radially and connecting the corresponding values into a polygon. Their appeal lies in the intuitive mapping between polygon shape and the underlying data, making them popular for visual comparisons at small dimensionalities. However, because features typically have no intrinsic circular order, the geometry of the radar chart -- and thus human interpretation -- depends greatly on how the axes are arranged. 

This dependence introduces a subtle but critical source of distortion. When adjacent features differ sharply in magnitude, the resulting polygon may exhibit exaggerated spikes or valleys, or even collapse inward when high and low values alternate. These artifacts can mislead viewers about the similarity or dissimilarity between profiles, creating illusory structure where none exists. Unlike time series or ordinal data, most radar chart features are unordered, meaning that any apparent trends or patterns can arise solely from arbitrary axis placement.

The interpretive consequences of axis ordering are well documented. Human vision tends to treat polygon shape as a meaningful pattern (\textit{shape-driven perception}), so abrupt transitions between neighboring features are perceived as instability or imbalance.  In multi-profile comparisons, poor ordering can also exaggerate or obscure genuine contrasts, reducing the chart’s communicative value.

In practical settings -- such as exploratory data analysis, advanced model explanation pipelines, or market analytics -- radar charts are often used to compare a small number of selected features across multiple profiles. In such contexts, an ordering that causes polygon collapse or excessive jaggedness is unappealing at best, and completely misleading at worst, as it may conceal structure that another permutation would reveal, or, conversely, artificially suggest structure where none exists.The challenge, then, is to determine a feature ordering that preserves geometric interpretability and visual balance across all profiles being compared.

Motivated by this problem, we introduce a deterministic, objective-driven reordering method that seeks feature permutations minimizing abrupt changes between consecutive axes. The proposed approach is data-driven, requires no manual intervention, transparent, and interpretable: it optimizes for smoothness across profiles without altering the traditional radar geometry. Rather than imposing semantic meaning on the order of features, our method aims to minimize visual distortion and enhance perceptual clarity by making the radar chart’s structure reflect the underlying relationships more faithfully.

\section{Related Work: How Our Approach Fits in the Landscape}

Several recent contributions address radar-chart limitations from different angles. We summarize the most relevant ones and explain where our method complements or differs from them.

\subsection{OrigamiPlot and area-invariant visual transforms}
Duan \emph{et al.} (2023) \cite{Duan2023} proposed the \emph{origami plot}, a visualization technique designed to improve the interpretability of radar-like visualizations by modifying the geometry so that the \textit{area of the connected region is invariant to attribute ordering}. While elegant, this approach alters the plot geometry itself. Our method, by contrast, preserves the traditional radar layout but optimizes the ordering to minimize spikiness and improve visual balance across profiles. Thus, OrigamiPlot removes the dependence by design, whereas we find the optimal permutation under the conventional encoding.

\subsection{Learning to order: Versatile Ordering Network (VON)}
The Versatile Ordering Network (VON) \cite{Yu2024} uses a neural and reinforcement-learning framework to learn orderings according to arbitrary differentiable metrics. This approach is powerful when many instances must be solved and amortized inference is beneficial. However, it is heuristic and requires training. Our method, in contrast, requires no prior training, is deterministic,  globally optimal and fast for small to moderate dimensionalities ($p$), offering an interpretable closed-form objective.

\subsection{Our contribution}
We contribute an explicit, transparent, and easily explainable lexicographic minimax criterion for radar chart ordering that:
\begin{itemize}
\item first minimizes the mean adjacent difference across all profiles (capturing overall smoothness),  
\item then uses the maximum jump as a tie-breaker to prevent extreme spikes,  
\item produces globally optimal, reproducible orderings for small to moderate $p$, and  
\item complements geometry-based or learned approaches by improving the \emph{existing} radar plot rather than replacing it.
\end{itemize}

\section{Problem Formulation}

Consider $m$ profiles $P_1, \dots, P_m$ each with $p$ normalized features,
\[
\mathbf{v}_j = [v_{j,1}, v_{j,2}, \dots, v_{j,p}] \quad \text{for } j = 1, \dots, m.
\]
Let $\pi$ denote a permutation of $\{1, \dots, p\}$ representing a feature order. For profile $j$, define the circular adjacent differences as
\[
\Delta \mathbf{v}_j^\pi = \big[|v_{j,\pi(2)} - v_{j,\pi(1)}|, \dots, |v_{j,\pi(p)} - v_{j,\pi(p-1)}|, |v_{j,\pi(1)} - v_{j,\pi(p)}|\big].
\]
We define two spikiness measures for profile $j$:
\[
\text{mean jump: } M_j(\pi) = \text{mean}(\Delta \mathbf{v}_j^\pi), \qquad
\text{max jump: } X_j(\pi) = \max(\Delta \mathbf{v}_j^\pi).
\]

The joint objective uses a lexicographic criterion:
\[
\pi^* = \arg\min_\pi \Big( \max_j M_j(\pi), \max_j X_j(\pi) \Big),
\]
which first prioritizes overall smoothness (mean jump) and then resolves ties by the worst single jump (max jump). This ordering preserves more global information than using the max first and then the mean for tie breaks, avoiding premature disqualification of otherwise smooth configurations.

\section{Proposed Strategy}

For small to moderate $p$, we evaluate all $p!$ permutations and select the permutation that minimizes the lexicographic pair $(\max_j M_j(\pi), \max_j X_j(\pi))$.  The rationale is that:

\begin{itemize}
\item By using a minimax framework we ensure that spikiness is reduced across all $j$ profiles simultaneously. Then, exhaustive evaluation guarantees that the chosen ordering is globally optimal, avoiding suboptimal arrangements that heuristics might produce.  In other words, this framework guarantees that spikiness is uniformly reduced across all profiles to the maximum extent possible given the data to be plotted. 
\item Using the mean first focuses on the global shape of the polygon, capturing overall smoothness.  
\item Using the max second prevents extreme spikes without discarding globally smooth solutions.  
\item Lexicographic ordering keeps more information about any given ordering than max-first: a good ordering with a slightly larger max jump but otherwise smooth is not unfairly rejected.  
\item Human perception is sensitive both to overall smoothness and local spikes; this strategy balances these perceptual factors.  
\end{itemize}

\subsection{The minimax framework and its application to radar charts}

The lexicographic minimax principle extends the classical minimax framework, a cornerstone of decision theory and robust optimization. In its standard form, the minimax criterion seeks the decision rule or configuration that minimizes the maximum possible loss across all scenarios or agents under consideration. This approach -- formally introduced by von Neumann \cite{vonneumann1928} and later generalized by Sion \cite{Sion1958} -- captures the notion of robustness: ensuring acceptable performance even under the most adverse conditions. Its influence extends widely, appearing in fields such as game theory, load balancing, scheduling, risk management, stress distribution and robust control.

In the context of feature ordering in radar charts, each candidate permutation of features can be viewed as a decision, and each profile as an ``agent" incurring a cost that reflects its geometric irregularity. The cost associated with a permutation is determined by local discontinuities between adjacent features -- abrupt jumps that visually manifest as spikes or distortions. The minimax objective, therefore, seeks the permutation that minimizes the largest such discontinuity across all profiles, ensuring that no profile exhibits a disproportionately irregular shape. This embodies a fairness principle akin to those used in multi-agent optimization, where the goal is to minimize the worst-off agent’s burden.

The lexicographic extension further refines this objective by introducing secondary criteria that act as tie-breakers among equally optimal minimax solutions. Specifically, we evaluate permutations first by their mean adjacent jump (a measure of overall smoothness) and second by their maximum jump (a measure of local extremity).  As introduced previously, this precise ordering of criteria is deliberate: by prioritizing the mean first, we preserve more information about the global distribution of feature differences before applying the more selective max operator. The mean captures aggregate smoothness across the entire profile set, while the max isolates the most pronounced local discontinuity. 

Applying the max operator first, in contrast, can lead to premature information loss:  focusing only on the worst local difference inherently disregards the broader distribution of feature differences, potentially overlooking permutations that achieve better overall smoothness despite a single, slightly larger jump.  By deferring the max operation, the lexicographic formulation balances global and local smoothness, yielding orderings that are both visually coherent and quantitatively robust. This structure mirrors the broader logic of robust optimization, where primary objectives capture typical performance and secondary objectives ensure resilience to worst-case deviations.

In summary, the proposed lexicographic minimax framework combines robustness (via the minimax component) with perceptual balance (via the mean-first criterion). This ensures that radar chart orderings remain stable, interpretable, and fair across profiles -- preventing the dominance of any single extreme while preserving the broader geometric structure that supports meaningful visual comparison.

\subsection{Complexity and practical recommendations}

The search space of all possible feature orderings grows factorially with the number of axes $p$. Exhaustive search is trivial for $p \le 7$ (5,040 permutations) and remains feasible up to roughly $p=10$ ($10! \approx 3.6$ million). Beyond $p \approx 12$, the number of possible orderings exceeds $479$ million, making exhaustive enumeration computationally prohibitive (with a single thread, that is). In such cases, heuristics (e.g. stochastic permutation selection) or learned orderings (e.g., clustering-based or neural approaches such as VON) become more practical.  

This combinatorial explosion is not unique to our formulation: it is directly analogous to the \emph{Traveling Salesman Problem} (TSP), one of the canonical NP-hard problems in combinatorial optimization \cite{GareyJohnson1979}. In TSP, a salesman must visit $p$ cities exactly once and return to the origin, minimizing total travel distance. Similarly, our problem seeks a cyclic permutation of $p$ features that minimizes a global cost function -- the spikiness criterion -- across multiple profiles. Both involve finding an optimal ordering over a complete graph under a symmetric cost function. Since in our case we must minimize a cost over multiple agents (or ``routes",  in the context of TSP) simultaneously, the complexity of this problem is at least that of the TSP.

No polynomial-time algorithm is known for the TSP, and the same complexity implications extend to our feature-ordering problem: unless P=NP, an efficient exact solution is unlikely to exist. Hence, the factorial computational cost of our exhaustive approach is theoretically justified given the global optimality it guarantees for up to moderate $p$. 

Conveniently, this theoretical limitation aligns well with the practical use of radar charts. Human interpretability declines sharply beyond about ten axes, as perceptual clutter and angular crowding obscure meaningful structure. Therefore, in the relevant range ($p \le 10$), our exhaustive minimax approach remains computationally feasible (often in practice, e.g. for $p \leq 7$, very inexpensive) while ensuring globally optimal and perceptually balanced orderings. For higher dimensions, approximate methods -- such as stochastic permutation selection, greedy insertion, simulated annealing, or reinforcement-based ordering networks -- could provide scalable alternatives without significant loss in visual quality.

\section{Implementation and Example}
\begin{figure}[h!]
\centering
\begin{minipage}{0.49\textwidth}
  \centering
  \includegraphics[width=\linewidth]{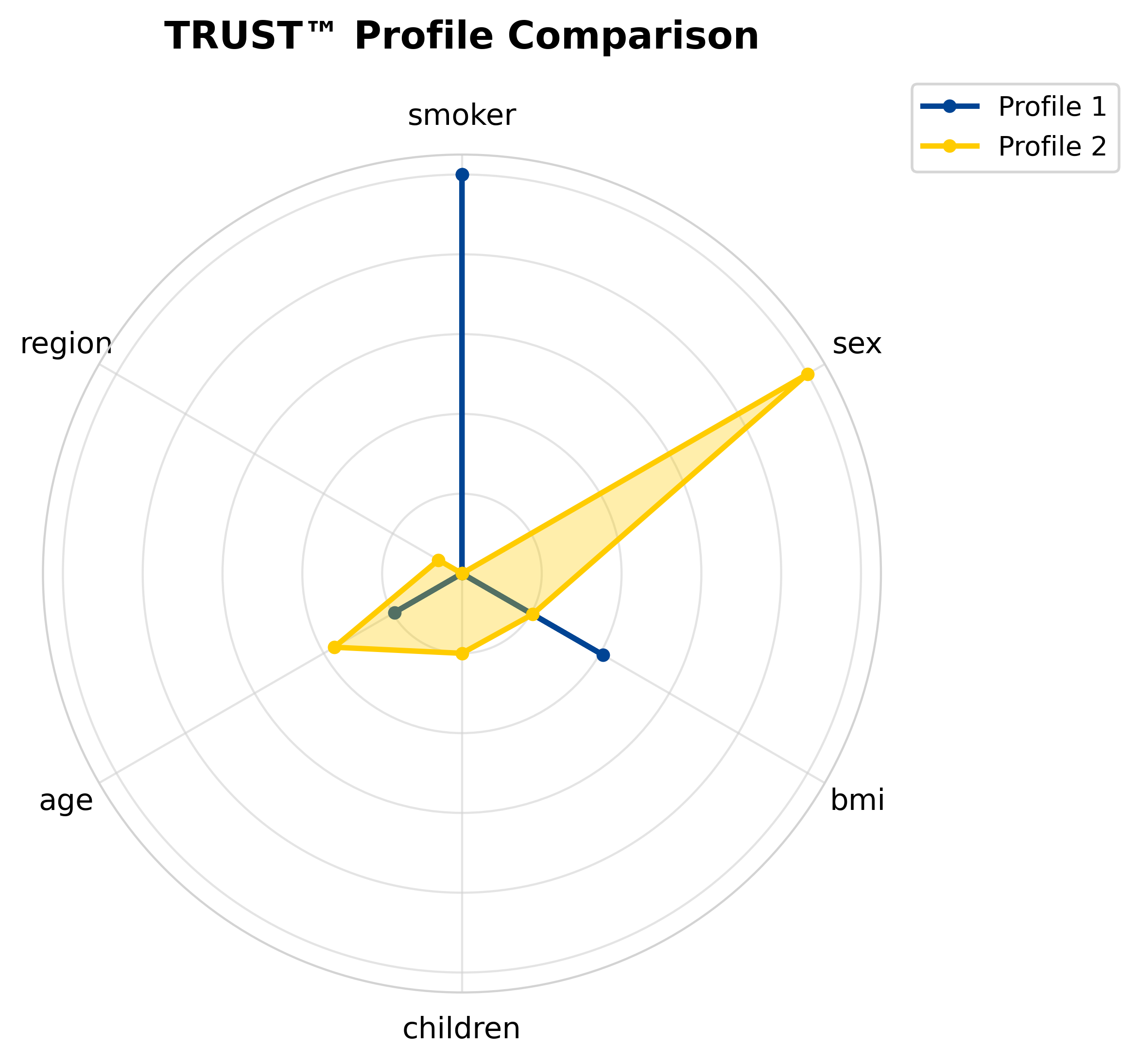}
  \\[3pt] 
  \textit{Before feature reordering}
\end{minipage}\hfill
\begin{minipage}{0.49\textwidth}
  \centering
  \includegraphics[width=\linewidth]{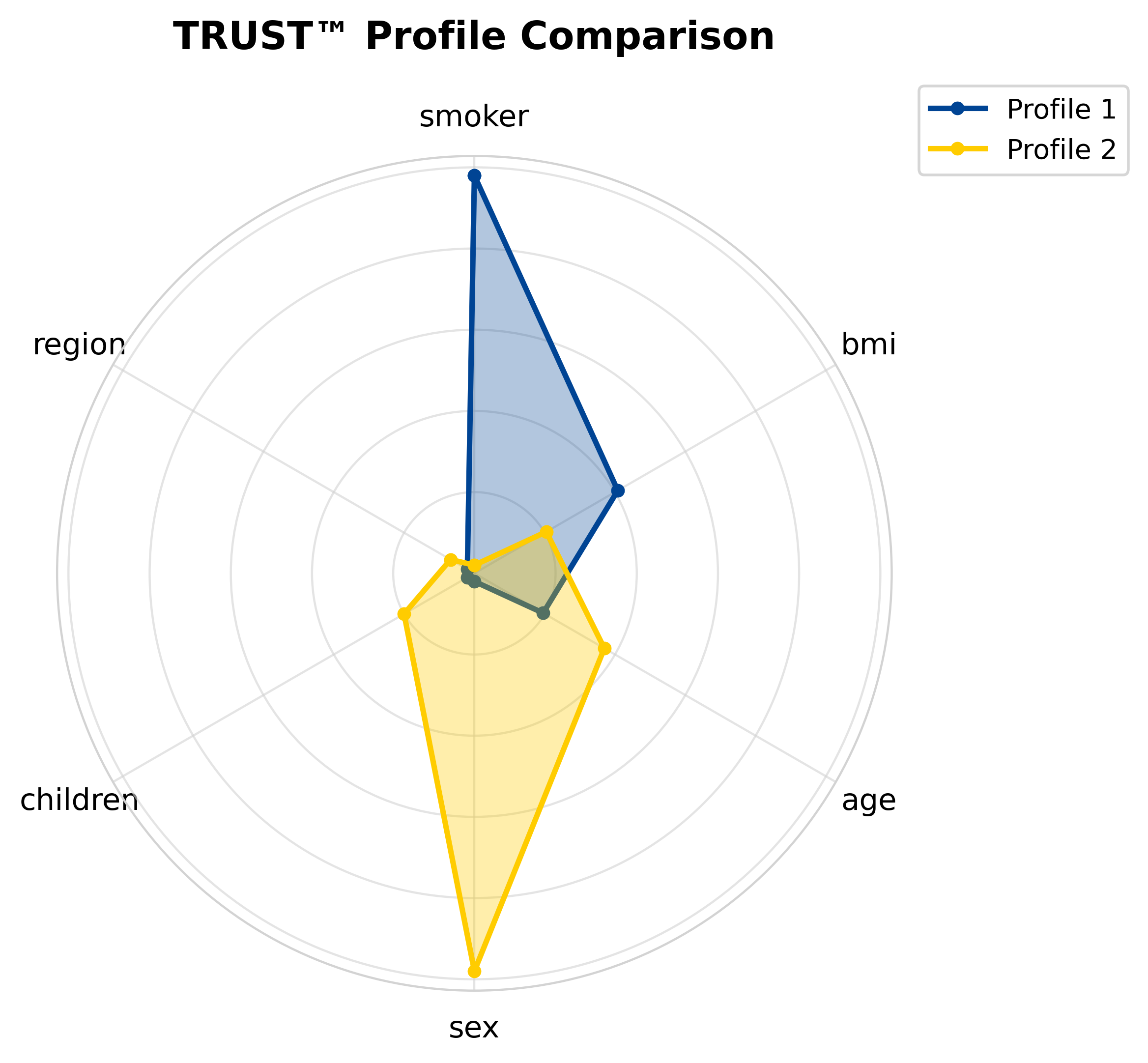}
  \\[3pt]
  \textit{After feature reordering}
\end{minipage}
\caption{Radar charts before (left) and after (right) reordering. The optimized ordering yields a smoother and more balanced shape.}
\label{fig:radar_example}
\end{figure}

We implemented the algorithm in Python using NumPy and Matplotlib. Small $\varepsilon$ offsets avoid degenerate zero-area polygons when normalized values hit boundaries, ensuring that even in the worst-case scenario profile areas don't vanish. Figure~\ref{fig:radar_example} illustrates a two-profile example with $p=6$ features,  showing substantial improvement in polygon smoothness after optimization. The categorical features `smoker', `sex' and `region' are target-encoded prior to being scaled just as the rest of variables. This is one way to adapt radar charts for categorical features. Feature scaling is necessary for two reasons: to have a common scale e.g. 0-1 for all features to be plotted, and related, to prevent from the different scales of features to arbitrarily influence the perception of similarity or dissimilarity between profiles.

\section{Limitations and Future Work}

The proposed lexicographic minimax approach for optimizing radar chart feature ordering, while effective for small to moderate feature sets, has certain limitations that warrant consideration. The reliance on exhaustive search, which evaluates all $p!$ permutations, results in factorial computational complexity with respect to the number of features $p$. Although this is feasible for moderate dimensions (e.g., $p \leq 10$), it becomes impractical for larger feature sets, where the combinatorial explosion renders exhaustive enumeration computationally prohibitive. To address this, future work could explore heuristic or approximate methods, such as stochastic permutation selection or greedy algorithms, to extend the approach’s scalability while maintaining near-optimal performance.

Additionally, the current framework employs mean and maximum jump measures to quantify spikiness, but alternative metrics, such as variance or percentile-based jump measures, could be investigated to capture different aspects of visual smoothness. The metric-agnostic nature of the proposed method facilitates such exploration, allowing for adaptation to specific visualization contexts. For scenarios involving many profiles, aggregating spikiness beyond the maximum jump -- potentially through percentile-based measures -- could enhance perceptual fairness by better balancing the visual impact across profiles. This would be particularly valuable in applications where diverse profile sets are compared simultaneously.

Finally, the lexicographic ordering principle itself offers opportunities for generalization. By incorporating weighted criteria or application-specific priorities, the framework could be tailored to emphasize particular perceptual or analytical goals, such as prioritizing certain features or aligning with domain-specific interpretability needs. Future research could also validate the perceptual benefits of optimized orderings through user studies, ensuring that the method not only improves quantitative metrics but also enhances human comprehension in practical settings.

\section{Conclusion}

This work introduces a principled, deterministic, and interpretable optimization strategy for reordering features in radar charts, specifically designed to enhance multi-profile comparisons. By employing a lexicographic minimax framework that prioritizes the mean adjacent jump to ensure overall polygon smoothness and uses the maximum jump as a tie-breaker to mitigate extreme spikes, the proposed method produces visually balanced and perceptually coherent radar charts. This approach effectively prevents visual collapse in overlapping profiles, thereby preserving the integrity of the data representation and facilitating more accurate interpretation of multivariate relationships. Unlike geometric transformations such as OrigamiPlot, which modify the radar chart’s structure, our method retains the traditional radar layout while significantly improving its interpretability and aesthetic quality. The transparency and global optimality of the algorithm, particularly for small to moderate feature sets, make it a practical tool for applications in exploratory data analysis, model explanation pipelines, and comparative visualization tasks. By addressing the critical issue of feature-ordering-induced distortion, this work contributes to the broader goal of designing robust and user-centric visualization tools, with potential for further refinement through scalable heuristics and perceptual validation studies.

\section*{Code Availability}
The method is implemented in the  \texttt{compare()} method included in the \texttt{trust-free} package, a free Python implementation of the regression TRUST™ algorithm \cite{Dorador2025} which, among multiple explanation tools,  it allows  multi-faceted profile comparisons that include a radar chart constructed with the data-driven optimized feature ordering that we have discussed in the present work.  See at \url{https://github.com/adc-trust-ai/trust-free} for the latest developments.


\begin{thebibliography}{99}

\bibitem{BenTal2009}
A.~Ben-Tal, L.~El Ghaoui, A.~Nemirovski, \emph{Robust Optimization}. Princeton University Press, 2009.

\bibitem{Dorador2025}
A.~Dorador,  ``TRUST: Transparent, Robust and Ultra-Sparse Trees",  arXiv: 2506.15791, 2025.

\bibitem{Duan2023}
R.~Duan, J.~Tong, A.~J. Sutton, D.~A. Asch, H.~Chu, C.~H. Schmid, Y.~Chen, ``Origami plot: a novel multivariate data visualization tool that improves radar chart",  \emph{Journal of Clinical Epidemiology}, 2023.

\bibitem{GareyJohnson1979}
M.~R. Garey and D.~S. Johnson, \emph{Computers and Intractability: A Guide to the Theory of NP-Completeness}. W. H. Freeman, 1979.

\bibitem{Sion1958}
M.~Sion, ``On general minimax theorems",  \emph{Pacific Journal of Mathematics}, vol. 8, pp. 171–176, 1958.

\bibitem{vonneumann1928}
J.~von Neumann, ``Zur Theorie der Gesellschaftsspiele", \emph{Mathematische Annalen}, vol. 100(1), pp. 295–320, 1928.

\bibitem{Yu2024}
Z.~Yu, W.~Zhang, S.~Pan, J.~Tao, ``Versatile Ordering Network: An Attention-based Neural Network for Ordering Across Scales and Quality Metrics",  arXiv:2412.12759, 2024.


\end{thebibliography}
\end{document}